# LUMPED ELEMENT MODELING OF NONLINEAR PROPERTIES OF HIGH TEMPERATURE SUPERCONDUCTORS IN A DIELECTRIC RESONATOR

### Dimitri Ledenyov*, Janina Mazierska*#, Greg Allen&, and Mohan Jacob*

***Abstract:*** *Microwave properties of High Temperature Superconductors (HTS) exhibit dependence on RF power levels. Effects of nonlinear phenomena causing this dependence can be modelled using lumped element circuits. Several RLC circuits to model the surface impedance of HTS were investigated (with equations expressing transmitted power derived) and assessed. Linear, quadratic and exponential dependences of surface resistance $R_S$ and inductance $L_S$ on the RF power for the model chosen as the best have been examined and used in simulations of electromagnetic responses of superconducting films in a Hakki-Coleman dielectric resonator.*

## 1. INTRODUCTION

Small losses of High Temperature Superconducting materials have resulted in applications of $YBa_2Cu_3O_{7-\delta}$ and $Tl_2CaBa_2Cu_2O_8$ films in microwave planar filters and in dielectric resonators. The surface impedance $Z_S=R_S+jX_S$ is the parameter describing microwave properties of HTS materials, where $R_S$ is the surface resistance and $X_S$ is the surface reactance, where $X_S=\omega L_S$, $L_S=\mu_0\omega\lambda$ is the surface inductance, $\mu_0$ is magnetic permeability of free space, $\omega$ is angular frequency and $\lambda$ is the penetration depth of a superconductor. An increase of the surface resistance $R_S$ and surface reactance $X_S$ with microwave power levels [1, 2] sets essential limitations to the application of HTS films in emerging device technologies. Despite the fact that the understanding of the physical phenomena behind the nonlinear effect has significantly advanced in recent years, the underlying mechanisms causing nonlinear behavior of $Z_S$ of HTS films are not yet fully understood [3-7]. The increase of the surface resistance of HTS films with RF power levels (as illustrated in Fig. 1) is attributed to one or more of the following phenomena: thermal effects, weak links, non-homogeneities, non-equilibrium excitation, unconventional pairing and others [9-13], each of which may occur under certain conditions. The dimensionless parameter *r*, defined as $\Delta X_S/\Delta R_S$, was proposed in [14] and as $\Delta R_S/\Delta X_S$ in [15], to differentiate mechanisms responsible for the nonlinear effect at varying power levels.

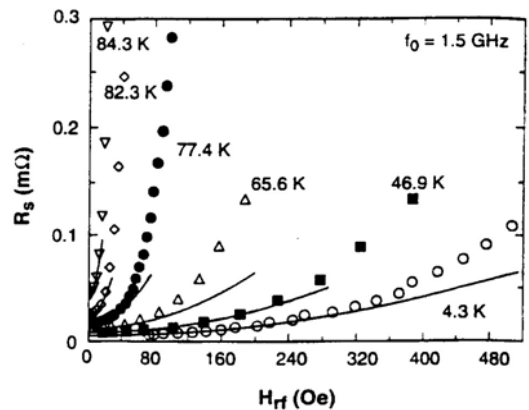

Fig. 1. $R_S$ vs $H_{RF}$ for a stripline resonator at differing temperatures after [8].

The various mechanisms affecting the nonlinear properties of High Temperature Superconductors each is said to have characteristic *r* values. Not only are the magnitudes of the r values are different but also the frequency and temperature dependence have characteristic signatures. In this context, *r* values can basically be used to differentiate among these mechanisms. For example, the weakly coupled grains model where the nonlinearity arises from the nonlinear inductance of the weak links, has the estimated r values lying around $10^{-3}$ [15]. The intrinsic nonlinearity mechanism [16] due to large transport currents from breaking of Cooper pairs into quasiparticles, is estimated to have the r values approximately $10^{-2}$ [15]. The uniform microwave heating mechanism, based on the fact that superconducting films have finite surface resistance and therefore absorb microwave power resulting in increase of their temperature, has the *r* values corresponding to 0.01-0.02 [15].

Another possible mechanism is creation of Josephson vortices in weak links (grain boundaries) by large microwave magnetic field. The nonlinearity arises from the dependence of vortex concentration on the $H_{rf}$, where the dissipation of the microwave energy is due to the viscous drag of the vortices. This model developed by Halbritter [14] predicts almost temperature and frequency independent r of value $\leq 1$.

---
*Electrical and Computer Engineering, &School of Information Technology, James Cook University, Townsville, Australia,
#Massey University, Institute of Information Sciences and Technology, Palmerston North, New Zealand,
email: j.mazierska@ieee.org

Large microwave magnetic field may cause vortices penetrate also the bulk, not just the grain boundaries, leading to the periodically change of the magnetic flux through the sample. When decreasing the field the vortices leave the grains, but their motion lags after the field due to viscous forces and hysteresis occurs. The value of *r* for the purely hysteretic losses is in the range of 1 [15] and depends only on the geometric form of the hysteresis curve.

In a local heating of weak links phenomenon high microwave currents may switch some weak links and areas near them to the normal state what leads to nonlinear effects. The order of the r parameter for the local heating was found to be high, $r > 100$ [15].

Differing values of r are believed to correspond to different mechanisms of nonlinearity. Hence a possibility exists of tracking and describing nonlinear properties of superconducting films based on the value of the parameter r exhibited by a particular HTS film under given $H_{rf}$. Therefore, a precise modeling of the nonlinear behavior of HTS materials may allow a prediction of performances of HTS based circuits and devices. Creation of a simple lumped element model representing electromagnetic behavior of a superconductor and identification of its parameters from basic measurements would enable easier analysis and design of HTS circuits.

To model the nonlinear electromagnetic response of a dielectric cavity with a HTS sample a simple parallel RLC circuit (Fig. 2(a)) was proposed in [17], with the microwave power expressed as [17, 18]:

$$P(\omega) = \frac{I^2}{2} \frac{R(P)}{1 + R(P)^2 (\omega C - \frac{1}{\omega L(P)})^2}$$

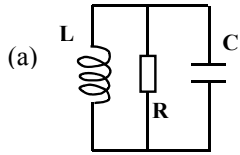
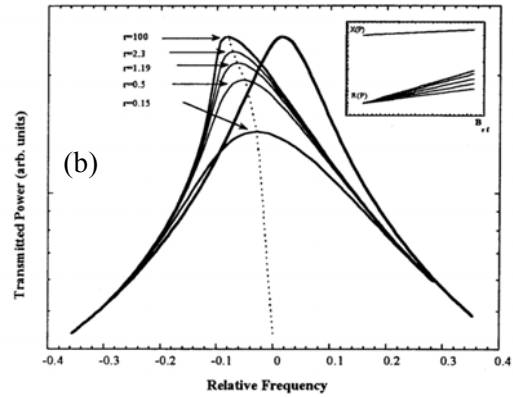

Fig. 2. (a) - Parallel RLC model, (b) - Simulated transmitted power vs frequency after [17]

The simple model of [17] with linear dependence of $R_s$ on $B_{rf}$ enabled simulations of transmitted power for varying values of the parameter r as shown in Fig. 2(b). However this model is not able to represent sufficiently behaviour of HTS films and a dielectric resonator. A more comprehensive approach for modelling of superconductors embedded in a dielectric resonator using RLC lumped element circuits is the aim of this paper.

## 2. MODELLING OF HTS FILMS IN A DIELECTRIC RESONATOR USING LUMPED ELEMENT CIRCUITS

Several RLC circuits can be used for modelling of an electromagnetic response of a dielectric resonator with HTS films. A dielectric resonator can be represented as either a parallel or a series RLC circuit, however for the parallel circuit, the higher resistance R, the higher quality factor, Q, of the network, hence a representation of the resonator as a series circuit is more appropriate for our case.

A superconductor is typically represented using the two-fluid model [19] consisting of a parallel combination of an inductor and a resistor [20], where the inductive part represents superconducting electrons and the resistive part - normal electrons in HTS materials. There are three possible equivalent circuits based on the two fluid model, not shown here due to lack of space. The model believed to represent the HTS in the best way is shown in Fig. 3. The superconductor is modeled by the inductance $L_{ss}$ representing the superconducting current flowing with no resistance, in parallel the normal current path modeled by inductance $L_{sn}$ and resistance $R_s$. The additional inductance $L_{sn}$ relates to the magnetic flux created by the magnetic field that still penetrates into the superconducting film even at 0K. Both inductances $L_{sn}$ and $L_{ss}$ have relatively small values as compared to the inductance L of the resonator. Hence the resonant frequency of the dielectric resonator is mostly determined by L, with a very little contribution from $L_{ss}$ and $L_{sn}$. A derived equation describing the transmitted RF power for the model shown in Fig.3 is given below:

$$P = \frac{1}{2} \frac{\omega^2 C^2 V^2 (L_{SS}^2 \omega^2 R_S + R R_S^2 + R\omega^2 L_{SS}^2 + 2R\omega^2 L_{SS} L_{Sn} + R\omega^2 L_{Sn}^2)}{2\omega^2 L_{SS} L_{Sn} + R^2 \omega^4 C^2 L_{SS}^2 + R_S^2 + \omega^2 L_{SS}^2 + \omega^2 L_{Sn}^2 + 2\omega^6 LC^2 L_{Sn}^2 L_{SS} + \omega^4 L_{SS}^2 C^2 R_S^2 +}$$

$$\overline{+ 2\omega^4 LC^2 R_S^2 L_{SS} + 2\omega^6 L^2 C^2 L_{SS} L_{Sn} + 2\omega^6 LC^2 L_{SS}^2 L_{Sn} + 2R\omega^4 C^2 L_{SS}^2 R_S + R^2 \omega^4 C^2 L_{Sn}^2}$$

$$\overline{+ \omega^4 L^2 C^2 R_S^2 - 4\omega^4 L_{SS} L_{Sn} C L_{Sn} + 2R^2 \omega^4 C^2 L_{SS} L_{Sn} - 2\omega^4 L_{SS}^2 LC - 2\omega^4 L_{SS}^2 C L_{Sn} - 2\omega^4 L_{Sn}^2 LC}$$

$$\overline{- 2\omega^4 L_{Sn}^2 L_{SS} C + R^2 \omega^2 C^2 R_S^2 + \omega^6 L^2 C^2 L_{SS}^2 + \omega^6 L^2 C^2 L_{Sn}^2 + \omega^6 L_{SS}^2 C^2 L_{Sn}^2 - 2R_S^2 \omega^2 LC - 2R_S^2 \omega^2 L_{SS} C}$$

(2)

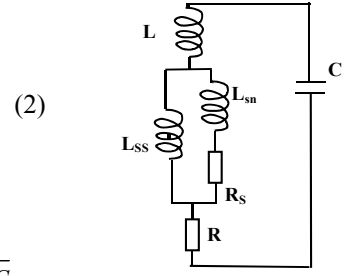

Fig. 3. Proposed Model

The equation (2) was used to simulate the nonlinear behaviour of High Temperature Superconductors when installed in a 10GHz Hakki Coleman dielectric resonator. Three types of power dependences of the model parameters $R_S$, $L_{SS}$ and $L_{Sn}$ as described in Section 3 were investigated.

## 3. POWER DEPENDENCES OF MODEL PARAMETERS $R_S$, $L_{SS}$ and $L_{Sn}$

To model the nonlinear behaviour of HTS materials the elements $R_{ss}$, $L_{ss}$, and $L_{sn}$ of the equivalent circuit of Fig. 3 need to be expressed as a function of transmitted RF power. There have been various mathematical dependencies proposed in terms of linear [17], [8], quadratic and exponential [21] dependencies on $H_{RF}$. In this paper three types of RF power dependences have been used to define $R_s$, $L_{ss}$, and $L_{sn}$. The expressions for each of the dependences are presented below:

| LINEAR | QUADRATIC | EXPONENTIAL |
|---|---|---|
| $R_S = R_{S0}(1 + \rho P(i-1))$; | $R_S = R_{S0}(1 + \rho_1 P(i-1) + \rho_2 P^2(i-1))$; | $R_S = R_{S0}(1 + a\,exp(bP(i-1)))$; |
| $L_{SS} = L_{SS0}(1 + lP(i-1))$; | $L_{SS} = L_{SS0}(1 + l_1 P(i-1) + l_2 P^2(i-1))$; | $L_{SS} = L_{SS0}(1 + c\,exp(dP(i-1)))$; |
| $L_{Sn} = L_{Sn0}(1 + lP(i-1))$; | $L_{Sn} = L_{Sn0}(1 + l_1 P(i-1) + l_2 P^2(i-1))$; | $L_{Sn} = L_{Sn0}(1 + c\,exp(dP(i-1)))$; |

(3)

where, $\rho$, $\rho_1$, $\rho_2$, $l$, $l_1$, $l_2$, $a$, $b$, $c$, and $d$ are fitting constants.

## 4. SIMULATED POWER RESPONSES USING THE PROPOSED MODEL

Simulations of the transmitted RF power of the Hakki-Coleman Dielectric Resonator of resonant frequency of 10GHz containing $YBa_2Cu_3O_7$ superconducting films using the model of Fig. 3 have been performed using a program written in Matlab. The following parameters have been assumed in the simulations:
Hakki-Coleman Dielectric Resonator: $R=3\times10^{-4}\Omega$, $C=10^{-12}F$, $L=2.5\times10^{-10}H$,
HTS films: $L_{ss}=L_{sn}=5\times10^{-13}H$ ($\lambda=4\times10^{-7}m$), $R_s=5\times10^{-4}\Omega$, $\rho=5.1\times10^{-4}$, $l=2.77\times10^{-5}$, $\rho_1=1.1\times10^{-3}$, $l_1=2.77\times10^{-5}$, $\rho_2=3.2\times10^{-7}$, $l_2=2\times10^{-8}$, $a=2.57\times10^{-1}$, $b=2.21\times10^{-3}$, $c=2.36\times10^{-2}$, $d=8\times10^{-4}$, $f_0=10GHz$.
The parameters $R_S$ and $L_S$ typical for $YBa_2Cu_3O_7$ at frequency of 10GHz and temperature of 77K have been assumed. Three power dependences as described by (3) were used in the simulations. Simulated power responses assuming an ideal and three power dependences are presented in Figs. 4. The ideal curve illustrates the case where $R_s$, $L_{ss}$, and $R_{sn}$ do not depend on RF power. As expected the exponential dependence resulted in the strongest nonlinear effect obtained. Computed transmitted power for the exponential dependence of $R_s$, $L_{ss}$, and $L_{sn}$ is given in Fig. 5 for the parameter r varying from 0.09 to 0.7. Fig. 6 and 7 illustrate results of simulations for the linear power dependence with fitting parameter for either $R_S$ of $L_{SS}$ (and $L_{Sn}$) varied.

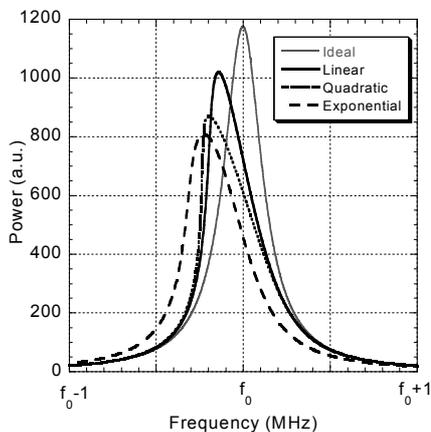

Fig. 4. $P_{RF}$ of HTS in DR for three power dependencies

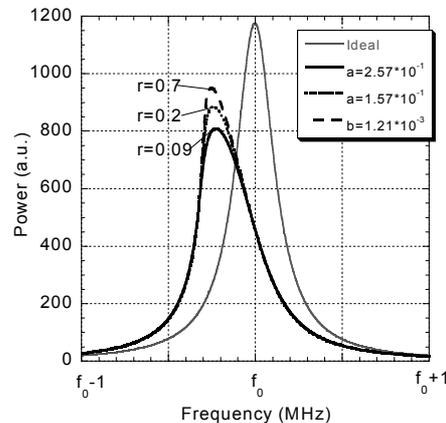

Fig. 5. Simulated $P_{RF}$ for the exponential dependence

As expected obtained results show that varying the parameter ρ affects the amplitude of the responses only, while changing the parameters l of $L_{SS}$ and $L_{Sn}$ tilts the resonance curves towards either higher of lower frequencies.

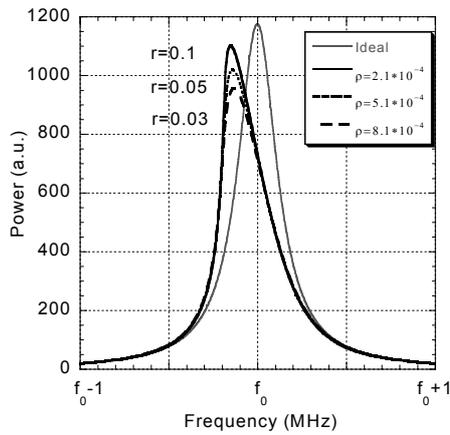 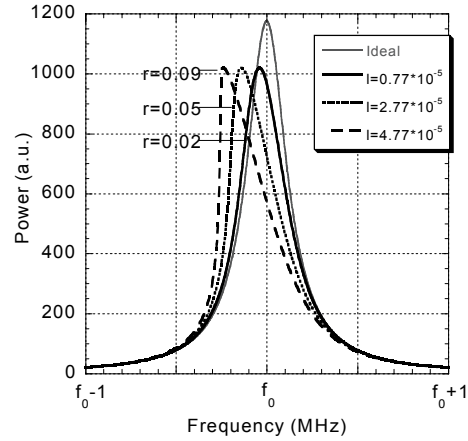

Fig. 6. Simulated $P_{RF}$ for the linear dependence with ρ varied     Fig. 7. Simulated $P_{RF}$ for the linear dependence with l varied

In performed simulations only one fitting parameter was changed at a time and all other characteristic system values were kept unchanged. Values of the coefficient ΔX/ΔR were varied from 0.02 to 1 in the computations; covering the case of intrinsic nonlinearity, nonlinearity due to weak links, Josephson vortices in weak links and vortices in bulk material.

## 5. CONCLUSION

Three lumped element models of a microwave resonator incorporating HTS thin films have been examined. A series resonant circuit representing the Hakki-Coleman dielectric resonator combined with a modified two fluid representation of HTS materials has proved to model microwave properties in the most suitable way. Performed simulations using the linear, quadratic and exponential dependence of $R_S$, $L_{SS}$ and $L_{Sn}$ of the RF power has shown the latter to give the strongest simulated nonlinear effect. The created model can represent comprehensively nonlinear properties of superconducting films in the dielectric resonator and will be used to predict EM response of HTS planar resonators.

**Acknowledgements**: This work was partly funded by the AC Large grant A00105170. The first author acknowledges the James Cook University PhD Scholarship and the MTT-IEEE Graduate Fellowship Award.